\documentclass[aps,
	       prl,
	       nofootinbib,
	       nobibnotes,notitlepage,
	       ]{revtex4-1} 

\usepackage{amssymb}
\usepackage{amsmath}
\usepackage{graphicx,subfigure}
\usepackage{longtable}
\usepackage{verbatim}
\usepackage{amsfonts}
\usepackage{hyperref}
\usepackage{xcolor}
\usepackage[normalem]{ulem}

\newcommand{\be}{\begin{equation}}
\newcommand{\ee}{\end{equation}}
\newcommand{\bea}{\begin{eqnarray}}
\newcommand{\eea}{\end{eqnarray}}
\newcommand{\beq}{\begin{equation}}
\newcommand{\eeq}{\end{equation}}
\def\beqa{\begin{eqnarray}}
  \def\eeqa{\end{eqnarray}}
\newcommand{\bv}{\left(\begin{array}{c}}
\newcommand{\ev}{\end{array}\right)}

\def\lsim{\mathrel{\rlap{\lower4pt\hbox{\hskip1pt$\sim$}}
    \raise1pt\hbox{$<$}}}	  
\def\gsim{\mathrel{\rlap{\lower4pt\hbox{\hskip1pt$\sim$}}
    \raise1pt\hbox{$>$}}}	  

\newcommand{\nn}{\nonumber}

\newcommand{\gev}{\ensuremath{\,}\mbox{GeV}}
\newcommand{\GeV}{\,\mbox{GeV}}

\newcommand{\mupi}{\mu_\pi^2}
\newcommand{\mug}{\mu_G^2}
\newcommand{\rd}{\rho_D^3}
\newcommand{\rls}{\rho_{LS}^3}

\begin{document}


\title{Three loop calculations and inclusive $V_{cb}$}
\author{Marzia Bordone}
\email{marzia.bordone@to.infn.it}
\author{Bernat Capdevila}
\email{bernat.capdevilasoler@unito.it}
\author{Paolo Gambino}
\email{paolo.gambino@unito.it}

\affiliation{
Dipartimento di Fisica, Universit\`a di Torino \& INFN, Sezione di Torino, I-10125 Torino, Italy}
\vspace*{1cm}

\begin{abstract}
We discuss the impact of the recent $\mathcal{O}(\alpha_s^3)$ calculations of the semileptonic width of the $b$ quark and of the relation between pole and kinetic heavy quark masses by Fael {\it et al.}\ on the inclusive determination of $|V_{cb}|$. The most notable effect 
is
a reduction of the uncertainty. Our final result is $|V_{cb}|=42.16(51)\, 10^{-3}$.

\end{abstract}

\maketitle

\section{Introduction}
The purpose of this note is to study the implications of the recent $\mathcal{O}(\alpha_s^3)$ calculations
by Fael, Sch\"onwald and Steinhauser \cite{Fael:2020tow,Fael:2020iea,Fael:2020njb}
on the determination of the Cabibbo-Kobayashi-Maskawa matrix element $|V_{cb}|$ from inclusive semileptonic $B$ decays, see Refs.~\cite{Alberti:2014yda,Gambino:2016jkc} for the most recent results. As is well-known,
the  values of $|V_{cb}|$ determined from
inclusive semileptonic $B$ decays and from $\bar B\to D^*\ell\bar\nu$ have differed
for a long time and, despite significant experimental and theoretical efforts, the situation remains quite confusing.
Recent accounts of the {\it $V_{cb}$ puzzle} can be found in Refs.~\cite{Gambino:2019sif,Bordone:2019vic,Gambino:2020jvv,Jaiswal:2020wer}. The latest lattice calculations \cite{Bazavov:2021bax,Harrison:2021tol,Kaneko:2019vkx,Martinelli:2021onb}, which for the first time explore the $\bar B\to D^*\ell\bar\nu$ form factors at non-zero recoil, have not clarified the issue and a preliminary comparison of these results shows  a few discrepancies \cite{Kaneko:talk}.
\vskip 2mm

The third order perturbative correction to the $b\to c \ell\bar\nu$ decay width computed  in Ref.~\cite{Fael:2020tow} and partially checked in Ref.~\cite{Czakon:2021ybq} represents  a fundamental step to improve the precision in the extraction of $|V_{cb}|$ from
inclusive $B$ decays. Indeed, perturbative corrections are sizeable -- they reduce the semileptonic rate by over 10\% -- and provide the dominant theoretical uncertainty. In the following we will
study the impact of the $\mathcal{O}(\alpha_s^3)$ corrections on the central value and uncertainty of $|V_{cb}|$.
\vskip 2mm

The other relevant three-loop calculation in our analysis is that of Refs.~\cite{Fael:2020iea,Fael:2020njb}, which concerns the relation between the pole (or $\overline{\rm MS}$)  and the kinetic masses of a heavy quark.  This calculation allows us to convert recent high-precision determinations of the $b$ quark  $\overline{\rm MS}$ mass \cite{Bazavov:2018omf,Aoki:2019cca} into the kinetic scheme \cite{Bigi:1996si} with an uncertainty of about 15 MeV, or 50\% less than the previous two-loop conversion \cite{Czarnecki:1997sz,Gambino:2011cq}. We will investigate the effect of such high-precision input in the global fit to the semileptonic moments and in the extraction of $|V_{cb}|$.
We should add that Ref.~\cite{Fael:2020njb} also computed the
charm mass effects at $\mathcal{O}(\alpha_s^2, \alpha_s^3)$ in the kinetic scheme relations.
We will also include $\mathcal{O}(\alpha_s \rd/m_b^3)$ effects from Ref.~\cite{Mannel:2019qel} in the semileptonic width and slightly update the global fit to the semileptonic moments.

\section{The total semileptonic width}
Our starting point is the Operator Product Expansion (OPE)
for the total semileptonic width (see Ref.~\cite{Gambino:2015ima} for a complete list of references):
\be
\Gamma_{sl}=\Gamma_0 f(\rho) \Big[1+ a_1 a_s+ a_2 a_s^2+ a_3 a_s^3-
\left(\frac12  -p_1  a_s \right)
\frac{\mu_\pi^2}{m_b^2}   + \left(g_0 +g_1 a_s\right)\frac{\mu_G^2(m_b)}{m_b^2}  +
 d_0 \frac{\rho_D^3}{m_b^3} -g_0 \frac{\rho_{LS}^3}{m_b^3} +\dots
\Big] \label{Gammasl}
\ee
where  $\Gamma_0= A_{ew}|V_{cb}|^2 G_F^2 m_b^{kin}(\mu)^5 /192 \pi^3$, $f(\rho)=1\!-\!8\rho\!+\!8\rho^3\! -\!\rho^4 \!-\!12\rho^2 \ln\rho$, $a_s=\alpha_s^{(4)}(\mu_b)/\pi$ is the strong coupling in the $\overline{\rm MS}$ scheme with 4 active quark flavours, $\rho=(\overline m_c(\mu_c)/m_b^{kin}(\mu))^2$ is the squared ratio of the $\overline{\rm MS}$ charm mass at the scale $\mu_c$,  $\overline m_c(\mu_c)$, and of the $b$ quark kinetic mass with a cutoff $\mu\sim1\,$GeV, $m_b^{kin}(\mu)$. $A_{ew}\simeq 1.014$ is the leading electroweak correction. The parameters  $\mu_\pi^2, \rho^3_D, $ etc.\
are nonperturbative expectation values of local operators in the $B$ meson defined in the kinetic scheme with cutoff $\mu$. They are generally extracted from a fit to central moments
of the lepton energy and of the hadronic invariant mass distributions in semileptonic $B$ decays \cite{Alberti:2014yda,Gambino:2016jkc}, for which the same contributions as in Eq.~\eqref{Gammasl} are included, with the exception of the $\mathcal{O}(\alpha_s^3)$ corrections which are available only for the width.
\vskip 2mm

The coefficients in Eq.~\eqref{Gammasl} depend on three unphysical scales: the scale of the $\overline{\rm MS}$ strong coupling constant $\mu_b$, that of the $\overline{\rm MS}$ charm
mass $\mu_c$, and the wilsonian cutoff $\mu$ employed in the kinetic scheme definition of the $b$ mass and of the OPE matrix elements. We choose  the $\overline{\rm MS}$ scheme for the charm mass because all high-precision determinations of this mass are expressed in this scheme and we prefer to escape uncertainties related to the scheme conversion; in the following we choose 1.6 GeV$\lsim\mu_c\lsim 3$ GeV, avoiding scales which are either too low or too high to provide a good convergence of the perturbative series.  The kinetic scheme \cite{Bigi:1996si,Czarnecki:1997sz,Fael:2020njb} provides a short-distance, renormalon-free definition of $m_b$ and of the
OPE parameters by introducing a hard cutoff $\mu$ to factor out the infrared contributions from the perturbative calculation. The cutoff $\mu$ should  ideally  satisfy $\Lambda_{\rm QCD}\ll \mu\ll m_b$; in the following we will vary it in the range 0.7--1.3\,GeV. Finally,
the scale of the strong coupling constant will be varied in the range 2--8$\,$GeV.
  Table I of Ref.~\cite{Alberti:2014yda} shows the size of the various coefficients in Eq.~\eqref{Gammasl}  for a couple of typical scale-settings.
The third order coefficient $a_3$ is new and stems from the calculation of Ref.~\cite{Fael:2020tow}.  We reproduce the numerical results of Ref.~\cite{Fael:2020tow}
 for the coefficients $a_i$.
\vskip 2mm

As a first step in our analysis, we employ the results of  the default fit of Ref.~\cite{Alberti:2014yda} with $\mu_b=m_b^{kin}$, $\mu_c=3\,$GeV, and $\mu=1\,$GeV,
to extract $|V_{cb}|$ from Eq.~(\ref{Gammasl}). To this end, we employ the total semileptonic branching fraction obtained in the same fit and $\tau_B=1.579(5)\,$ps \cite{Amhis:2019ckw}. Notice that the 2014 default fit included a constraint on $\overline{m}_c(3\, \mathrm{GeV})$, but not on $m_b^{kin}$.
Small shifts have to be applied to the values of $m_b^{kin}$, $\mu_\pi^2$ and $\rho_D^3$ extracted from the fit in order to account for
missing two-loop charm mass effects in the  kinetic scheme definition adopted
in the 2014 fit. These effects have now been computed in Ref.~\cite{Fael:2020njb},
where it was found that they reduce to decoupling effects and that they can be taken into account
by expressing the kinetic scheme definitions in terms of  $\alpha_s^{(3)}$.
For $\mu_b=m_b^{kin}$ the shifts amount to $+4\,$MeV, $-0.003\,$GeV$^2$, $-0.002\,$GeV$^3$, respectively.
We adopt the PDG value for $\alpha_s^{(5)}(M_Z)=0.1179(10)$, from which we get
   $\alpha_s^{(4)}(4.557 {\rm GeV})=0.2182(36)$. We employ RunDec  \cite{Chetyrkin:2000yt} to compute the running of all relevant scale-dependent quantities.
 We eventually
 find $|V_{cb}|=42.49(44)_{th}(33)_{exp}\ 10^{-3}$, where the uncertainty refers to the inputs {\it only} and is split into an experimental and  a theoretical component. If we neglect the new three-loop result we recover the same $|V_{cb}|$ central value  as in \cite{Alberti:2014yda}.
The three loop correction therefore shifts   $|V_{cb}|$ by +0.6\%, well within the theoretical uncertainty of 1.3\% estimated in \cite{Alberti:2014yda}. The perturbative series is
\be
\Gamma_{sl}=\Gamma_0 f(\rho)\Big[ 
{ 0.9255- 0.1162_{\alpha_s} } - 
 { 0.0350_{ \alpha_s^2} }- 0.0097_{\alpha_s^3}\Big] = 
{ 0.5401}\, \Gamma_0,\label{eq:2}
\ee
where the first term differs from 1 because of the power corrections.
We can also repeat the same exercise evolving the value of $\overline{m}_c(3\,\mathrm{GeV})=0.987(13)\,$GeV from the fit to $\mu_c=2\,$GeV, which gives  $\overline{m}_c(2\,\mathrm{GeV})=1.091(14)\,$GeV and extract again $|V_{cb}|$ using $\mu_c=2\,$GeV in Eq.~\eqref{Gammasl}. We get $|V_{cb}|=42.59(44)_{th}(33)_{exp} \ 10^{-3}$
and
\be
\Gamma_{sl}=\Gamma_0 f(\rho)\Big[0.9258 - 0.0878_{\alpha_s}  - 0.0179_{ \alpha_s^2} - 0.0005_{\alpha_s^3}\Big]=  0.5374 \,\Gamma_0.\label{eq:3}
\ee
As  noted in Ref.~\cite{Fael:2020tow}, the better convergence of the perturbative expansion with $\mu_c=2\,$GeV,  already observed in Ref.~\cite{Gambino:2013rza}, carries on  at the three loop level, but the cancellations appear somewhat accidental. Since the physical scale of the decay is actually lower than $m_b$, we believe a more appropriate choice  for the scale of $\alpha_s^{(4)}$ is  $\mu_b=m_b^{kin}/2$, which with $\mu_c=2\,$GeV leads to $|V_{cb}|=42.59(44)_{th}(33)_{exp} \ 10^{-3}$ and 
\be
\Gamma_{sl}=\Gamma_0 f(\rho)\Big[{0.9255} - 0.1140_{\alpha_s}  - 0.0011_{ \alpha_s^2} +{ 0.0103}_{\alpha_s^3}\Big]={0.5381}\, \Gamma_0.\label{eq:4}
\ee
\vskip 2mm

We see from Eqs.~\eqref{eq:2}-\eqref{eq:4} that the typical size of the three-loop corrections is 1\% and that the perturbative series converge well at different values of the scales. A conservative estimate of the residual perturbative uncertainty on $\Gamma_{sl}$ is therefore 0.5\%,
but it is worth studying the scale dependence of the width in more detail.
In Fig.~1 we show the $\mu_b$ and $\mu_c$ dependence of Eq.~\eqref{Gammasl} at two and three loops, using the inputs of the 2014 default fit. The scale dependence is reduced by the inclusion of the three loop contribution by over a factor 2, and the red curves appear to be flatter than the blue ones. The region of minimal scale dependence is the one around $\mu_b\sim\mu_c\approx 2-3$ GeV. We also studied the dependence of the width on the kinetic scale in the range $0.7<\mu<1.3\gev$ for different values of $\mu_{b,c}$, finding similar results.
Defining $\Delta_{\mu_{b,c}}$ the maximum percentage deviation of the red solid lines in Fig.~1 and $\Delta_{\mu}$ accordingly, with $\mu_b=m_b^{kin}/2$ and $\mu_c=2\text{GeV}$, we get
\be
\Delta_{\mu_{b}}=
0.44\%\,, \quad 
\qquad
\Delta_{\mu_{c}}=
0.44\%\,, 
\qquad
\Delta_{\mu}=0.67\%\,.
\ee
Based on all this, we  conservatively estimate a residual perturbative uncertainty of  0.7\% in $\Gamma_{sl}$ and consequently of 0.35\% in $|V_{cb}|$ for our new default scenario, corresponding to  $\mu=1\gev$, $\mu_c=2\gev$ and $\mu_b=m_b^{kin}/2\simeq 2.3\gev$.

\begin{figure}[t]
\begin{center}
\includegraphics[height=5cm]{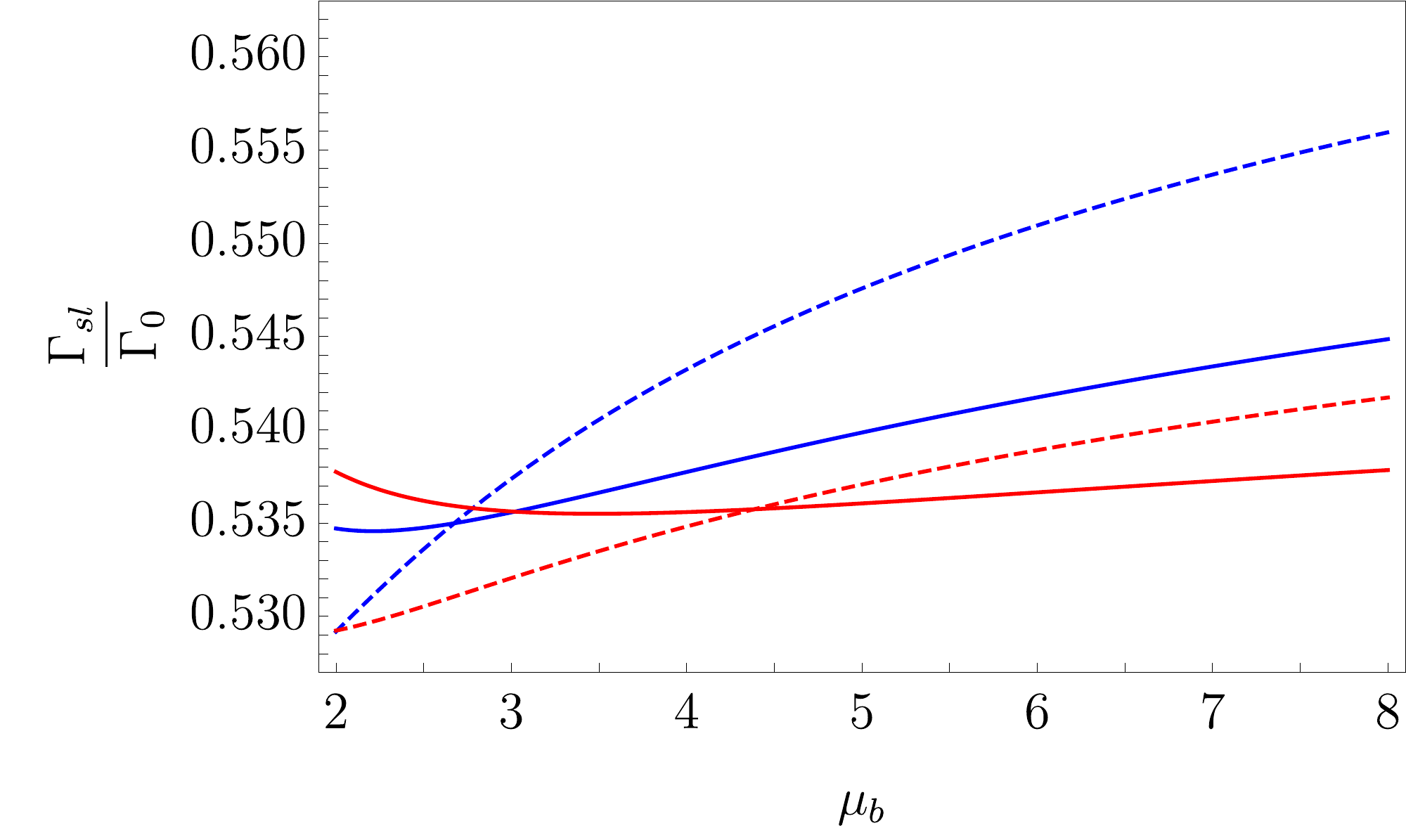}
\includegraphics[height=5cm]{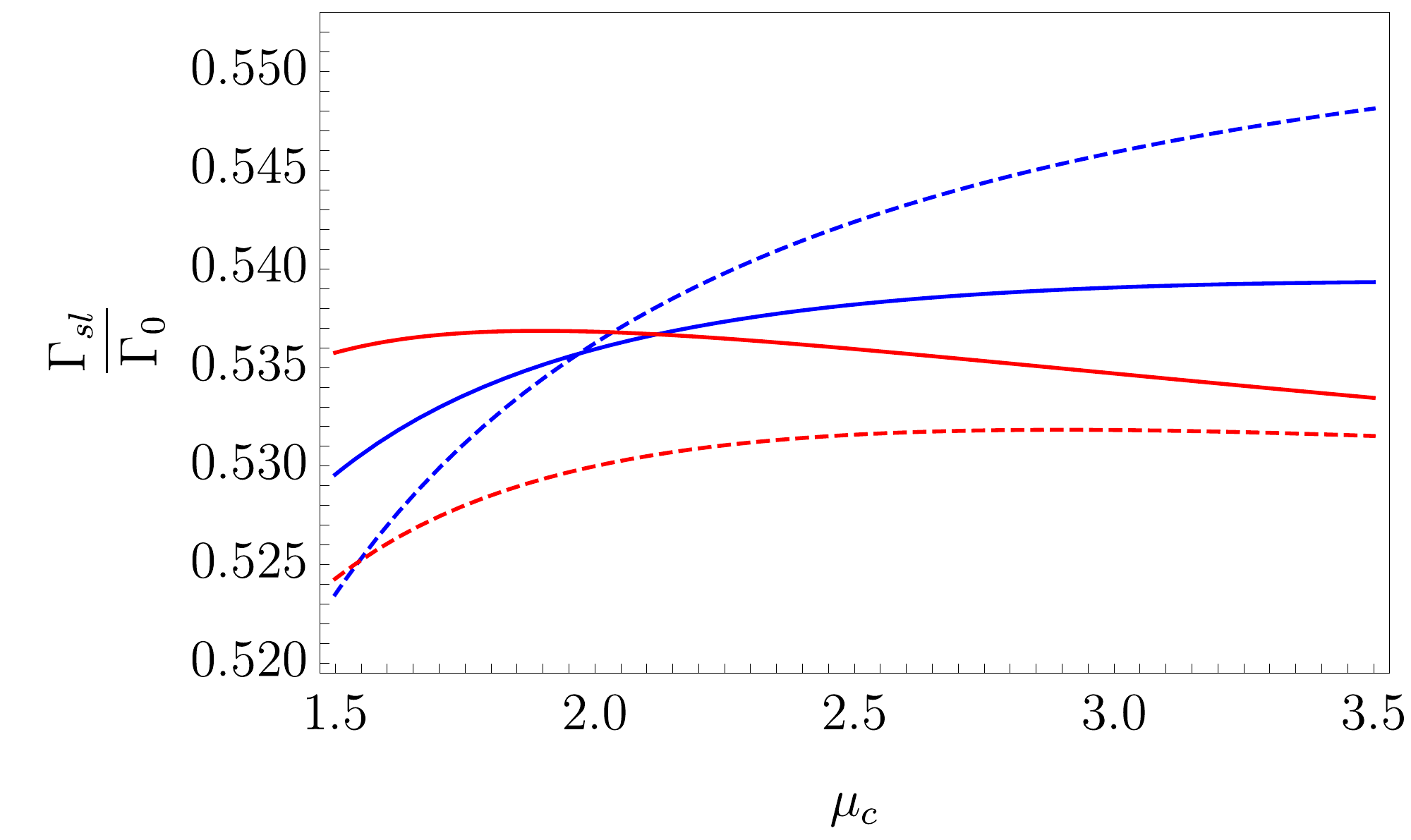}
\end{center}
\caption{ Scale dependence of $\Gamma_{sl}$ at fixed values of the inputs and $\mu^{kin}=1$GeV. Dashed (solid) lines represent the two (three) loop calculation. In the left  plot ($\mu_b$-dependence) the blue (red)
curves are at $\mu_c=3$(2)GeV; in the right plot ($\mu_c$-dependence) the blue(red) curves  $\mu_b=m_b^{kin}(m_b^{kin}/2)$. 
}\label{mu-dep}
\end{figure}
\vskip 2mm

Beside the purely perturbative contributions, there are various other sources of uncertainty in the calculation of the semileptonic width \cite{Benson:2003kp}, but the work done in the last few years has been fruitful. After the $\mathcal{O}(\alpha_s/m_b^2)$ corrections
\cite{Alberti:2013kxa,Mannel:2015jka},
 the  $\mathcal{O}(\alpha_s \rho_D^3/m_b^3)$ corrections to  $\Gamma_{sl}$ have been recently computed in Ref.~\cite{Mannel:2019qel} (the $\mathcal{O}(\alpha_s \rho_{LS}^3)$ corrections to  $\Gamma_{sl}$ follow from the $\mathcal{O}(\alpha_s\mu_G^2/m_b^2)$ and are tiny). They are expressed in terms of $m_b$ in the on-shell scheme and of $\overline{m}_c(m_b)$. After converting their result to the kinetic scheme and changing the scale of $\overline{m}_c$, we find that this new correction, together with all the terms of the same order  generated by the change of scheme, enhances the coefficient of $\rho_D^3$ by 8 to 18\%,  depending on the various scales. However, after the conversion to the kinetic scheme the $\mathcal{O}(\alpha_s\rd)$ terms generate
new $\mathcal{O}(\mu^3\alpha_s^2)$ and $\mathcal{O}(\mu^3\alpha_s^3)$ contributions that tend to compensate their effect. The resulting final shift on $|V_{cb}|$ is $ +0.05\%$ with $\mu_c=3\gev, \mu_b=m_b^{kin}$ and $+0.1\%$ for $\mu_c=2\gev, \mu_b=m_b^{kin}/2$, and we choose to neglect it in the following.
\vskip 2mm

After the  calculation of the $\mathcal{O}(\alpha_s\rho_D^3)$ contribution, the main residual
uncertainty in $\Gamma_{sl}$ is related to higher power corrections. The Wilson coefficients of the $\mathcal{O}(1/m_b^4, 1/m^5)$ contributions have been computed at the tree level \cite{Mannel:2010wj}  --- here the $\mathcal{O}(1/m^5)$ effects include $\mathcal{O}(1/m_b^3 m_c^2)$, sometimes referred to as Intrinsic Charm --- but little is known about the corresponding 27 matrix elements. The Lowest Lying State Approximation (LLSA) \cite{Mannel:2010wj} has been employed to estimate them and to guide the extension \cite{Gambino:2016jkc} of Ref.~\cite{Alberti:2014yda} to  $\mathcal{O}(1/m_b^4, 1/m^5)$. In the LLSA, the  $\mathcal{O}(1/m_b^4, 1/m^5)$ contributions increase  the width by about 1\%, but there is an important  interplay with the semileptonic fit: as shown in Ref.~\cite{Gambino:2016jkc}, the
$\mathcal{O}(1/m_b^4, 1/m^5)$
corrections to the moments and their uncertainties  modify the results of the fit in a subtle way and the final change in $\Gamma_{sl}$ is about +0.5\%, a result stable under changes of the LLSA assumptions \cite{Gambino:2016jkc}. We therefore expect the $\mathcal{O}(1/m_b^4, 1/m^5)$
corrections to decrease $|V_{cb}|$ by 0.25\% with respect to the default fit. Although the uncertainty attached to this value is mostly included in the theoretical uncertainty of the 2014 fit results, we may consider  an additional 0.3\% uncertainty for the width.
Further uncertainties stem from
unknown $\mathcal{O}(\alpha_s^2/m_b^2)$, and $\mathcal{O}(\alpha_s^2 \rho_{D}^3/m_b^3)$ corrections, but they are all likely to be at or below the 0.1\% level.
 The so-called Intrinsic Charm contributions, related to soft charm, lead to the $\mathcal O(1/m_b^3 m_c^2)$ corrections mentioned above, but also to terms of $\mathcal O(\alpha_s/m_b^3 m_c)$ which may contribute up to 0.5\%
to the width \cite{Bigi:2005bh}.  Finally, one expects 
quark-hadron duality  to break down at some point. Combining all the discussed sources of uncertainty, we estimate the total remaining uncertainty in  $\Gamma_{sl}$ to be 1.2\%.
\vskip 2mm

In the end, using the inputs of the 2014 default fit and setting $\mu_c=2\,\mathrm{GeV}$, $\mu_b=m_b^{kin}/2$ for the central value, we obtain
\be
|V_{cb}|_{2014}=42.49(44)_{th}(33)_{exp} (25)_\Gamma \,10^{-3}= 42.48(60)\,10^{-3}\label{Vcbfixed}
\ee
where the uncertainty due to $\Gamma_{sl}$ has been reduced by a factor 2 with respect to Ref.~\cite{Alberti:2014yda}.

\section{Updating the semileptonic fit}

Despite ongoing analyses  of the $q^2$ and $M_X$-moments at Belle and Belle II \cite{Abudinen:2020zwm,vanTonder:2021cyp}, no new experimental result on the semileptonic moments has been published since
the 2014 fit  \cite{Alberti:2014yda}. On the other hand, new lattice determinations of $m_b$ and $m_c$ have been presented, improving their precision by roughly a factor 2. We use the  FLAG 2019 averages \cite{Aoki:2019cca} with $N_f=2+1+1$ for $m_b$ and $m_c$,
\begin{align}
& \overline m_c(\rm 3 GeV)=0.988(7) \rm GeV,\nn\\
&  \overline m_b( \overline m_b)=4.198(12) \rm GeV,\label{eq:masses}
 \end{align}
which correspond to $\overline m_c(2\, \mathrm{GeV})=1.093(8)$ and
$m_b^{kin}({\rm 1 GeV})=4.565(19)$GeV, where for the latter
we have used option  B of \cite{Fael:2020njb} for the definition of $m_b^{kin}$.
We now repeat the 2014 default fit with both these constraints, slightly updating the
theoretical uncertainty estimates. In view of the small impact of the $\mathcal{O}(1/m_b^{4}, 1/m^5)$   and $\mathcal{O}(\alpha_s \rd)$ corrections discussed in the previous section, we
reduce the theoretical  uncertainties used in the fit to the moments with respect to Ref.~\cite{Alberti:2014yda}. In particular, we
consider a 20\%, instead of a 30\%, shift in $\rho_D^3$ and $\rho_{LS}^3$, and reduce to 4 MeV the safety shift in $m_{c,b}$. For all of the other settings and for the selection of experimental data
we follow Ref.~\cite{Alberti:2014yda}.
\vskip 2mm

While the central values of the fit are close to those of 2014, the uncertainty on $m_b^{kin}$ ($\overline{m}_c(3\gev)$) decreases from 20(12) to 12(7) MeV, and we get
$
|V_{cb}|=42.39(32)_{th}(32)_{exp}(25)_\Gamma\ 10^{-3}
$
with  $\chi^2_{min}/dof=0.46$. The very same fit performed with  $\mu_c=2\,$GeV and $\mu_b=m_b^{kin}/2$ gives
\be
|V_{cb}|=42.16(30)_{th}(32)_{exp}(25)_\Gamma\ 10^{-3} \label{default}
\ee
 with  $\chi^2_{min}/dof=0.47$ and we neglect the very small shift due to the $\mathcal O(\alpha_s \rd)$ correction to $\Gamma_{sl}$.
This is our new reference value and in Table I we display the complete results of this fit.

\begin{table}[t]
  \begin{center} \begin{tabular}{cccccccc}\hline
 $m_b^{kin}$ & $\overline m_c(2\rm GeV)$   &  $\mupi $ &$\rd$ &$\mug(m_b)$ & $\rls$  & ${\rm BR}_{c\ell\nu}$ & $10^3|V_{cb}|$ \\ \hline
     4.573 &1.092 & 0.477 & 0.185 & 0.306 & -0.130 & 10.66 & 42.16 \\
    0.012 & 0.008 & 0.056 & 0.031 & 0.050 & 0.092 & \ 0.15 & \ 0.51\\ \hline
   1
 & 0.307 & -0.141 & 0.047 & 0.612 & -0.196 & -0.064 & -0.420\\
  & 1 & 0.018 & -0.010 & -0.162 & 0.048 & 0.028&0.061\\
 &  & 1& 0.735 & -0.054 & 0.067 & 0.172 &0.429\\
  &  &  & 1 & -0.157 & -0.149 & 0.091& 0.299\\
  &  &  &  & 1 & 0.001 & 0.013& -0.225\\
  &  &  &  &  & 1 & -0.033 &-0.005\\
  &  &
 &  &  &  & 1 &    0.684\\
   &  &  &  &  &  & & 1    \\  \hline
   \end{tabular} \end{center}
       \caption{ \label{tab:2}
    Results of the updated fit in our default scenario ($\mu_c=2\gev, \mu_b=m_b^{kin}/2$).
All parameters are in $\GeV$ at the appropriate power and all, except $m_c$, in the kinetic scheme at $\mu=1\GeV$. The first and second rows give central values and uncertainties, the correlation matrix follows. }
\end{table}

\vskip 2mm

Let us now comment on the interplay between the fit to the moments and the use of Eq.~\eqref{Gammasl}. First, we observe that the fit to the moments is based on an $\mathcal{O}(\alpha_s^2)$ calculation \cite{Melnikov:2008qs,Pak:2008qt,Pak:2008cp,Biswas:2009rb,Gambino:2011cq} without $\mathcal{O}(\alpha_s\rho_D^3)$ contributions,
and that the lower precision in the calculation of the moments  with respect to the width inevitably affects the determination of $|V_{cb}|$. This is clearly visible in Eq.~\eqref{Vcbfixed}, where the theoretical component of the
error is larger than the residual theory error associated with the width.
However, only a small part of that uncertainty is related to the purely perturbative corrections,
which are relatively suppressed in some semileptonic moments but  sizeable in $\Gamma_{sl}$, as we have seen above. In other words, an $\mathcal{O}(\alpha_s^3)$ calculation of the moments is unlikely to improve the precision of the fit significantly, and
the inclusion of $\mathcal{O}(\alpha_s^3)$  corrections only in $\Gamma_{sl}$ is perfectly justified.
 On the other hand,  an $\mathcal{O}(\alpha_s/m_b^3)$ calculation of the moments can have an important
impact on the $|V_{cb}|$ determination. This is because the semileptonic moments, and the hadronic central moments in particular, are highly sensitive to the OPE parameters. Since
the power correction related to $\rd$ amounts to about 3\% percent in Eq.~\eqref{Gammasl}, an $\mathcal{O}(\alpha_s)$ shift on $\rd$ induced by perturbative corrections to  the moments can have a significant impact in the determination of $|V_{cb}|$.
Our estimates of the theoretical uncertainties take this into account. We also note that  a fit without theoretical errors is a very poor fit ($\chi^2/dof\sim 2$) with  $|V_{cb}|$ decreased by slightly less than $ 1\,\sigma$.
\vskip 2mm

An important problem of the semileptonic fit is the sensitivity to the ansatz employed for
the correlation among the theoretical uncertainties associated with the various observables \cite{Gambino:2013rza}. We have studied the dependence of the result of Eq.~(\ref{default})
 on the modelling of the theoretical correlations following Ref.~\cite{Gambino:2013rza} closely. Since the results shown above have been obtained using scenario {\bf D} from Ref.~\cite{Gambino:2013rza}
with $\Delta=0.25$GeV, we have repeated the fit with option {\bf B}, with option {\bf C}  using
various values of $\xi$, and with option {\bf D} for $\Delta$ in the range $0.15-0.30\,$GeV. The central values
for $|V_{cb}|$ vary between 42.05 $10^{-3}$ and 42.28 $10^{-3}$.   These results are very much in line
with Fig.~1 and Table 3 of Ref.~\cite{Gambino:2013rza} and therefore we do not add any uncertainty related to the theoretical correlations in Eq.~(\ref{default}).
\vskip 2mm

We have also performed a fit including $\mathcal{O}(1/m_b^4, 1/m^5)$ corrections, in analogy with Ref.~\cite{Gambino:2016jkc}, to check the consistency with our main result of Eq.~(\ref{default}).  We assign an error to the LLSA predictions and assume gaussian priors for all of the 27 dimension 7 and 8 matrix elements. The error is chosen as the maximum of either 60\% of the parameters value in the LLSA or $\Lambda_{LL}^n/2$ ($n = 4, 5$), with $\Lambda_{LL}=0.55\gev$, see Ref.~\cite{Gambino:2016jkc} for additional details. As already noticed in Ref.~\cite{Gambino:2016jkc}, higher power corrections tend to  decrease the value of $|V_{cb}|$. A fit performed with the same theory errors of Ref.~\cite{Gambino:2016jkc} and $\mu_c=2\,\mathrm{GeV}$ and $\mu_b=m_b^{kin}/2$ leads to $|V_{cb}|=42.00(32)_{th}(32)_{exp}(25)_\Gamma\ 10^{-3}=42.00(53)\ 10^{-3}$, which is consistent with Eq.~(\ref{default}). Following the discussion above, one could slightly reduce the theory uncertainties 
in this fit with the only consequence of a small reduction on the error of $|V_{cb}|$.
\vskip 2mm

Finally, repeating the reference fit of Table \ref{tab:2} without a constraint on $m_b$
we obtain an independent determination of $m_b^{kin}(1\gev)=4.579(16)\gev$, which 
translates into $\overline{m}_b(\overline{m}_b)=4.210(22)\gev$. This determination, which 
still relies on the lattice determination of $m_c$ reported in  (\ref{eq:masses}), is compatible with the FLAG $N_f=2+1+1$ average for  $\overline{m}_b(\overline{m}_b)$ and competitive with other current determinations of $m_b$.

\section{Discussion}
From a theoretical point of view, the reliability of the determination of $|V_{cb}|$ from inclusive semileptonic decays depends on our control of higher order effects. The new three-loop calculation of Ref.~\cite{Fael:2020tow} shows that higher order perturbative effects are
under control, and that they lie within the previously estimated uncertainties. This progress, together with the work done on higher power corrections \cite{Mannel:2010wj,Gambino:2016jkc} and on perturbative corrections to the Wilson coefficients of power suppressed operators \cite{Alberti:2013kxa,Mannel:2015jka,Mannel:2019qel}, led us to
estimate a residual theoretical error on $\Gamma_{sl}$ of about 1.2\%, and to slightly reduce the theoretical uncertainty in the fit to the moments.

Our final result is shown in Eq.~(\ref{default}). It is very close to previous determinations of
$|V_{cb}|$ \cite{Alberti:2014yda, Gambino:2016jkc}, but the total uncertainty is now 1.2\%, one
third smaller than in \cite{Alberti:2014yda}. This reduction of the uncertainty reflects a better control of higher order effects,  but it is also due to
improved determinations of the heavy quark masses.
 The dominant single component of the uncertainty in Eq.~(\ref{default}) is now related to the experimental determination of the moments and of the semileptonic branching fraction, which are expected to be improved at Belle II.
Future experimental analyses should also consider new observables beyond the traditional moments of the lepton energy and hadronic invariant mass distributions. For instance, the forward-backward asymmetry \cite{Turczyk:2016kjf} and the moments of the leptonic invariant mass ($q^2$) distribution would enhance the sensitivity to the OPE matrix elements and reduce the uncertainty on $|V_{cb}|$.
Because of reparametrisation invariance, the $q^2$-moments and $\Gamma_{sl}$ depend on a reduced number of OPE matrix elements   \cite{Fael:2018vsp}, so that a fit to the $q^2$-moments at $\mathcal{O}(1/m_b^4)$ involves only 8 parameters. This nice property allows for an independent check of the treatment of higher power corrections adopted in \cite{Gambino:2016jkc}, but it is unlikely to lead to a competitive determination of $|V_{cb}|$.
 The $q^2$-moments will also constrain the soft charm effects considered in \cite{Bigi:2005bh}.
As far as the current experimental analyses are concerned, there are various aspects that require closer scrutiny. We refer in particular to the subtraction of QED corrections made with PHOTOS~\cite{Golonka:2005pn}, to the impact of Coulomb interactions, to the contribution of $D^{**}$ states and to the correlations which play a crucial role in the fit.
\vskip 2mm

Finally, turning to ways in which theory can improve the inclusive determination of  $|V_{cb}|$,
we have already argued that  the most important missing higher order effects are probably the  $\mathcal{O}(\alpha_s/m_b^3)$ corrections to the moments.  Lattice QCD calculations
provide precise constraints on the heavy quark masses, see Eq.~\eqref{eq:masses}, which are going
to improve in the future, but  we now have  methods  to compute differential distributions and
their moments directly on the lattice  \cite{Gambino:2020crt}. While it is still unclear
whether  a
determination of $\Gamma_{sl}$ competitive with Eq.~(\ref{Gammasl}) can be achieved at the physical $b$ mass,  these lattice calculations might be able to
enhance the predictive power of the OPE by accessing quantities which are inaccurate or beyond the reach of current experiments and are highly sensitive to the non-perturbative parameters.
The computation of meson masses at different quark mass values \cite{Bazavov:2018omf,Gambino:2017vkx}  can also provide useful information when the data are analysed in the  heavy quark expansion.
At the moment, however,  the reduction of the uncertainty  in Eq.~(\ref{default}) exacerbates the $V_{cb}$ puzzle, and calls for renewed efforts to solve an unwelcome anomaly,
impervious to New Physics explanations \cite{Jung:2018lfu,Crivellin:2014zpa}.

\vspace{3mm}

{\bf Acknowledgements}
 We are grateful to Matteo Fael for useful communications concerning the results of Refs.~\cite{Fael:2020tow,Fael:2020iea,Fael:2020njb}.
This work is supported   by the Italian Ministry of Research (MIUR) under grant PRIN  20172LNEEZ.

\bibliography{VcbInclusive.bib}

\begin{thebibliography}{43}%
\makeatletter
\providecommand \@ifxundefined [1]{%
 \@ifx{#1\undefined}
}%
\providecommand \@ifnum [1]{%
 \ifnum #1\expandafter \@firstoftwo
 \else \expandafter \@secondoftwo
 \fi
}%
\providecommand \@ifx [1]{%
 \ifx #1\expandafter \@firstoftwo
 \else \expandafter \@secondoftwo
 \fi
}%
\providecommand \natexlab [1]{#1}%
\providecommand \enquote  [1]{``#1''}%
\providecommand \bibnamefont  [1]{#1}%
\providecommand \bibfnamefont [1]{#1}%
\providecommand \citenamefont [1]{#1}%
\providecommand \href@noop [0]{\@secondoftwo}%
\providecommand \href [0]{\begingroup \@sanitize@url \@href}%
\providecommand \@href[1]{\@@startlink{#1}\@@href}%
\providecommand \@@href[1]{\endgroup#1\@@endlink}%
\providecommand \@sanitize@url [0]{\catcode `\\12\catcode `\$12\catcode
  `\&12\catcode `\#12\catcode `\^12\catcode `\_12\catcode `\%12\relax}%
\providecommand \@@startlink[1]{}%
\providecommand \@@endlink[0]{}%
\providecommand \url  [0]{\begingroup\@sanitize@url \@url }%
\providecommand \@url [1]{\endgroup\@href {#1}{\urlprefix }}%
\providecommand \urlprefix  [0]{URL }%
\providecommand \Eprint [0]{\href }%
\providecommand \doibase [0]{http://dx.doi.org/}%
\providecommand \selectlanguage [0]{\@gobble}%
\providecommand \bibinfo  [0]{\@secondoftwo}%
\providecommand \bibfield  [0]{\@secondoftwo}%
\providecommand \translation [1]{[#1]}%
\providecommand \BibitemOpen [0]{}%
\providecommand \bibitemStop [0]{}%
\providecommand \bibitemNoStop [0]{.\EOS\space}%
\providecommand \EOS [0]{\spacefactor3000\relax}%
\providecommand \BibitemShut  [1]{\csname bibitem#1\endcsname}%
\let\auto@bib@innerbib\@empty
\bibitem [{\citenamefont {Fael}\ \emph
  {et~al.}(2021{\natexlab{a}})\citenamefont {Fael}, \citenamefont
  {Sch\"onwald},\ and\ \citenamefont {Steinhauser}}]{Fael:2020tow}%
  \BibitemOpen
  \bibfield  {author} {\bibinfo {author} {\bibfnamefont {M.}~\bibnamefont
  {Fael}}, \bibinfo {author} {\bibfnamefont {K.}~\bibnamefont {Sch\"onwald}}, \
  and\ \bibinfo {author} {\bibfnamefont {M.}~\bibnamefont {Steinhauser}},\
  }\href {\doibase 10.1103/PhysRevD.104.016003} {\bibfield  {journal} {\bibinfo
   {journal} {Phys. Rev. D}\ }\textbf {\bibinfo {volume} {104}},\ \bibinfo
  {pages} {016003} (\bibinfo {year} {2021}{\natexlab{a}})},\ \Eprint
  {http://arxiv.org/abs/2011.13654} {arXiv:2011.13654 [hep-ph]} \BibitemShut
  {NoStop}%
\bibitem [{\citenamefont {Fael}\ \emph {et~al.}(2020)\citenamefont {Fael},
  \citenamefont {Sch\"onwald},\ and\ \citenamefont
  {Steinhauser}}]{Fael:2020iea}%
  \BibitemOpen
  \bibfield  {author} {\bibinfo {author} {\bibfnamefont {M.}~\bibnamefont
  {Fael}}, \bibinfo {author} {\bibfnamefont {K.}~\bibnamefont {Sch\"onwald}}, \
  and\ \bibinfo {author} {\bibfnamefont {M.}~\bibnamefont {Steinhauser}},\
  }\href {\doibase 10.1103/PhysRevLett.125.052003} {\bibfield  {journal}
  {\bibinfo  {journal} {Phys. Rev. Lett.}\ }\textbf {\bibinfo {volume} {125}},\
  \bibinfo {pages} {052003} (\bibinfo {year} {2020})},\ \Eprint
  {http://arxiv.org/abs/2005.06487} {arXiv:2005.06487 [hep-ph]} \BibitemShut
  {NoStop}%
\bibitem [{\citenamefont {Fael}\ \emph
  {et~al.}(2021{\natexlab{b}})\citenamefont {Fael}, \citenamefont
  {Sch\"onwald},\ and\ \citenamefont {Steinhauser}}]{Fael:2020njb}%
  \BibitemOpen
  \bibfield  {author} {\bibinfo {author} {\bibfnamefont {M.}~\bibnamefont
  {Fael}}, \bibinfo {author} {\bibfnamefont {K.}~\bibnamefont {Sch\"onwald}}, \
  and\ \bibinfo {author} {\bibfnamefont {M.}~\bibnamefont {Steinhauser}},\
  }\href {\doibase 10.1103/PhysRevD.103.014005} {\bibfield  {journal} {\bibinfo
   {journal} {Phys. Rev. D}\ }\textbf {\bibinfo {volume} {103}},\ \bibinfo
  {pages} {014005} (\bibinfo {year} {2021}{\natexlab{b}})},\ \Eprint
  {http://arxiv.org/abs/2011.11655} {arXiv:2011.11655 [hep-ph]} \BibitemShut
  {NoStop}%
\bibitem [{\citenamefont {Alberti}\ \emph {et~al.}(2015)\citenamefont
  {Alberti}, \citenamefont {Gambino}, \citenamefont {Healey},\ and\
  \citenamefont {Nandi}}]{Alberti:2014yda}%
  \BibitemOpen
  \bibfield  {author} {\bibinfo {author} {\bibfnamefont {A.}~\bibnamefont
  {Alberti}}, \bibinfo {author} {\bibfnamefont {P.}~\bibnamefont {Gambino}},
  \bibinfo {author} {\bibfnamefont {K.~J.}\ \bibnamefont {Healey}}, \ and\
  \bibinfo {author} {\bibfnamefont {S.}~\bibnamefont {Nandi}},\ }\href
  {\doibase 10.1103/PhysRevLett.114.061802} {\bibfield  {journal} {\bibinfo
  {journal} {Phys. Rev. Lett.}\ }\textbf {\bibinfo {volume} {114}},\ \bibinfo
  {pages} {061802} (\bibinfo {year} {2015})},\ \Eprint
  {http://arxiv.org/abs/1411.6560} {arXiv:1411.6560 [hep-ph]} \BibitemShut
  {NoStop}%
\bibitem [{\citenamefont {Gambino}\ \emph {et~al.}(2016)\citenamefont
  {Gambino}, \citenamefont {Healey},\ and\ \citenamefont
  {Turczyk}}]{Gambino:2016jkc}%
  \BibitemOpen
  \bibfield  {author} {\bibinfo {author} {\bibfnamefont {P.}~\bibnamefont
  {Gambino}}, \bibinfo {author} {\bibfnamefont {K.~J.}\ \bibnamefont {Healey}},
  \ and\ \bibinfo {author} {\bibfnamefont {S.}~\bibnamefont {Turczyk}},\ }\href
  {\doibase 10.1016/j.physletb.2016.10.023} {\bibfield  {journal} {\bibinfo
  {journal} {Phys. Lett.}\ }\textbf {\bibinfo {volume} {B763}},\ \bibinfo
  {pages} {60} (\bibinfo {year} {2016})},\ \Eprint
  {http://arxiv.org/abs/1606.06174} {arXiv:1606.06174 [hep-ph]} \BibitemShut
  {NoStop}%
\bibitem [{\citenamefont {Gambino}\ \emph {et~al.}(2019)\citenamefont
  {Gambino}, \citenamefont {Jung},\ and\ \citenamefont
  {Schacht}}]{Gambino:2019sif}%
  \BibitemOpen
  \bibfield  {author} {\bibinfo {author} {\bibfnamefont {P.}~\bibnamefont
  {Gambino}}, \bibinfo {author} {\bibfnamefont {M.}~\bibnamefont {Jung}}, \
  and\ \bibinfo {author} {\bibfnamefont {S.}~\bibnamefont {Schacht}},\ }\href
  {\doibase 10.1016/j.physletb.2019.06.039} {\bibfield  {journal} {\bibinfo
  {journal} {Phys. Lett. B}\ }\textbf {\bibinfo {volume} {795}},\ \bibinfo
  {pages} {386} (\bibinfo {year} {2019})},\ \Eprint
  {http://arxiv.org/abs/1905.08209} {arXiv:1905.08209 [hep-ph]} \BibitemShut
  {NoStop}%
\bibitem [{\citenamefont {Bordone}\ \emph {et~al.}(2020)\citenamefont
  {Bordone}, \citenamefont {Jung},\ and\ \citenamefont {van
  Dyk}}]{Bordone:2019vic}%
  \BibitemOpen
  \bibfield  {author} {\bibinfo {author} {\bibfnamefont {M.}~\bibnamefont
  {Bordone}}, \bibinfo {author} {\bibfnamefont {M.}~\bibnamefont {Jung}}, \
  and\ \bibinfo {author} {\bibfnamefont {D.}~\bibnamefont {van Dyk}},\ }\href
  {\doibase 10.1140/epjc/s10052-020-7616-4} {\bibfield  {journal} {\bibinfo
  {journal} {Eur. Phys. J. C}\ }\textbf {\bibinfo {volume} {80}},\ \bibinfo
  {pages} {74} (\bibinfo {year} {2020})},\ \Eprint
  {http://arxiv.org/abs/1908.09398} {arXiv:1908.09398 [hep-ph]} \BibitemShut
  {NoStop}%
\bibitem [{\citenamefont {Gambino}\ \emph {et~al.}(2020)\citenamefont {Gambino}
  \emph {et~al.}}]{Gambino:2020jvv}%
  \BibitemOpen
  \bibfield  {author} {\bibinfo {author} {\bibfnamefont {P.}~\bibnamefont
  {Gambino}} \emph {et~al.},\ }\href {\doibase 10.1140/epjc/s10052-020-08490-x}
  {\bibfield  {journal} {\bibinfo  {journal} {Eur. Phys. J. C}\ }\textbf
  {\bibinfo {volume} {80}},\ \bibinfo {pages} {966} (\bibinfo {year} {2020})},\
  \Eprint {http://arxiv.org/abs/2006.07287} {arXiv:2006.07287 [hep-ph]}
  \BibitemShut {NoStop}%
\bibitem [{\citenamefont {Jaiswal}\ \emph {et~al.}(2020)\citenamefont
  {Jaiswal}, \citenamefont {Nandi},\ and\ \citenamefont
  {Patra}}]{Jaiswal:2020wer}%
  \BibitemOpen
  \bibfield  {author} {\bibinfo {author} {\bibfnamefont {S.}~\bibnamefont
  {Jaiswal}}, \bibinfo {author} {\bibfnamefont {S.}~\bibnamefont {Nandi}}, \
  and\ \bibinfo {author} {\bibfnamefont {S.~K.}\ \bibnamefont {Patra}},\ }\href
  {\doibase 10.1007/JHEP06(2020)165} {\bibfield  {journal} {\bibinfo  {journal}
  {JHEP}\ }\textbf {\bibinfo {volume} {06}},\ \bibinfo {pages} {165} (\bibinfo
  {year} {2020})},\ \Eprint {http://arxiv.org/abs/2002.05726} {arXiv:2002.05726
  [hep-ph]} \BibitemShut {NoStop}%
\bibitem [{\citenamefont {Bazavov}\ \emph {et~al.}(2021)\citenamefont {Bazavov}
  \emph {et~al.}}]{Bazavov:2021bax}%
  \BibitemOpen
  \bibfield  {author} {\bibinfo {author} {\bibfnamefont {A.}~\bibnamefont
  {Bazavov}} \emph {et~al.} (\bibinfo {collaboration} {Fermilab Lattice,
  MILC}),\ }\href@noop {} {\  (\bibinfo {year} {2021})},\ \Eprint
  {http://arxiv.org/abs/2105.14019} {arXiv:2105.14019 [hep-lat]} \BibitemShut
  {NoStop}%
\bibitem [{\citenamefont {Harrison}\ and\ \citenamefont
  {Davies}(2021)}]{Harrison:2021tol}%
  \BibitemOpen
  \bibfield  {author} {\bibinfo {author} {\bibfnamefont {J.}~\bibnamefont
  {Harrison}}\ and\ \bibinfo {author} {\bibfnamefont {C.~T.~H.}\ \bibnamefont
  {Davies}} (\bibinfo {collaboration} {LATTICE-HPQCD}),\ }\href@noop {} {\
  (\bibinfo {year} {2021})},\ \Eprint {http://arxiv.org/abs/2105.11433}
  {arXiv:2105.11433 [hep-lat]} \BibitemShut {NoStop}%
\bibitem [{\citenamefont {Kaneko}\ \emph {et~al.}(2019)\citenamefont {Kaneko}
  \emph {et~al.}}]{Kaneko:2019vkx}%
  \BibitemOpen
  \bibfield  {author} {\bibinfo {author} {\bibfnamefont {T.}~\bibnamefont
  {Kaneko}} \emph {et~al.} (\bibinfo {collaboration} {JLQCD}),\ }\href
  {\doibase 10.22323/1.363.0139} {\bibfield  {journal} {\bibinfo  {journal}
  {PoS}\ }\textbf {\bibinfo {volume} {LATTICE2019}},\ \bibinfo {pages} {139}
  (\bibinfo {year} {2019})},\ \Eprint {http://arxiv.org/abs/1912.11770}
  {arXiv:1912.11770 [hep-lat]} \BibitemShut {NoStop}%
\bibitem [{\citenamefont {Martinelli}\ \emph {et~al.}(2021)\citenamefont
  {Martinelli}, \citenamefont {Simula},\ and\ \citenamefont
  {Vittorio}}]{Martinelli:2021onb}%
  \BibitemOpen
  \bibfield  {author} {\bibinfo {author} {\bibfnamefont {G.}~\bibnamefont
  {Martinelli}}, \bibinfo {author} {\bibfnamefont {S.}~\bibnamefont {Simula}},
  \ and\ \bibinfo {author} {\bibfnamefont {L.}~\bibnamefont {Vittorio}},\
  }\href@noop {} {\  (\bibinfo {year} {2021})},\ \Eprint
  {http://arxiv.org/abs/2105.08674} {arXiv:2105.08674 [hep-ph]} \BibitemShut
  {NoStop}%
\bibitem [{\citenamefont {Kaneko}()}]{Kaneko:talk}%
  \BibitemOpen
  \bibfield  {author} {\bibinfo {author} {\bibfnamefont {T.}~\bibnamefont
  {Kaneko}} (\bibinfo {collaboration} {JLQCD}),\ }\href@noop {} {\bibinfo
  {journal} {talk at FPCP 2021}\ }\BibitemShut {NoStop}%
\bibitem [{\citenamefont {Czakon}\ \emph {et~al.}(2021)\citenamefont {Czakon},
  \citenamefont {Czarnecki},\ and\ \citenamefont {Dowling}}]{Czakon:2021ybq}%
  \BibitemOpen
\bibfield  {journal} {  }\bibfield  {author} {\bibinfo {author} {\bibfnamefont
  {M.}~\bibnamefont {Czakon}}, \bibinfo {author} {\bibfnamefont
  {A.}~\bibnamefont {Czarnecki}}, \ and\ \bibinfo {author} {\bibfnamefont
  {M.}~\bibnamefont {Dowling}},\ }\href@noop {} {\  (\bibinfo {year} {2021})},\
  \Eprint {http://arxiv.org/abs/2104.05804} {arXiv:2104.05804 [hep-ph]}
  \BibitemShut {NoStop}%
\bibitem [{\citenamefont {Bazavov}\ \emph {et~al.}(2018)\citenamefont {Bazavov}
  \emph {et~al.}}]{Bazavov:2018omf}%
  \BibitemOpen
  \bibfield  {author} {\bibinfo {author} {\bibfnamefont {A.}~\bibnamefont
  {Bazavov}} \emph {et~al.} (\bibinfo {collaboration} {Fermilab Lattice, MILC,
  TUMQCD}),\ }\href {\doibase 10.1103/PhysRevD.98.054517} {\bibfield  {journal}
  {\bibinfo  {journal} {Phys. Rev. D}\ }\textbf {\bibinfo {volume} {98}},\
  \bibinfo {pages} {054517} (\bibinfo {year} {2018})},\ \Eprint
  {http://arxiv.org/abs/1802.04248} {arXiv:1802.04248 [hep-lat]} \BibitemShut
  {NoStop}%
\bibitem [{\citenamefont {Aoki}\ \emph {et~al.}(2020)\citenamefont {Aoki} \emph
  {et~al.}}]{Aoki:2019cca}%
  \BibitemOpen
  \bibfield  {author} {\bibinfo {author} {\bibfnamefont {S.}~\bibnamefont
  {Aoki}} \emph {et~al.} (\bibinfo {collaboration} {Flavour Lattice Averaging
  Group}),\ }\href {\doibase 10.1140/epjc/s10052-019-7354-7} {\bibfield
  {journal} {\bibinfo  {journal} {Eur. Phys. J. C}\ }\textbf {\bibinfo {volume}
  {80}},\ \bibinfo {pages} {113} (\bibinfo {year} {2020})},\ \Eprint
  {http://arxiv.org/abs/1902.08191} {arXiv:1902.08191 [hep-lat]} \BibitemShut
  {NoStop}%
\bibitem [{\citenamefont {Bigi}\ \emph {et~al.}(1997)\citenamefont {Bigi},
  \citenamefont {Shifman}, \citenamefont {Uraltsev},\ and\ \citenamefont
  {Vainshtein}}]{Bigi:1996si}%
  \BibitemOpen
  \bibfield  {author} {\bibinfo {author} {\bibfnamefont {I.~I.~Y.}\
  \bibnamefont {Bigi}}, \bibinfo {author} {\bibfnamefont {M.~A.}\ \bibnamefont
  {Shifman}}, \bibinfo {author} {\bibfnamefont {N.}~\bibnamefont {Uraltsev}}, \
  and\ \bibinfo {author} {\bibfnamefont {A.~I.}\ \bibnamefont {Vainshtein}},\
  }\href {\doibase 10.1103/PhysRevD.56.4017} {\bibfield  {journal} {\bibinfo
  {journal} {Phys. Rev. D}\ }\textbf {\bibinfo {volume} {56}},\ \bibinfo
  {pages} {4017} (\bibinfo {year} {1997})},\ \Eprint
  {http://arxiv.org/abs/hep-ph/9704245} {arXiv:hep-ph/9704245} \BibitemShut
  {NoStop}%
\bibitem [{\citenamefont {Czarnecki}\ \emph {et~al.}(1998)\citenamefont
  {Czarnecki}, \citenamefont {Melnikov},\ and\ \citenamefont
  {Uraltsev}}]{Czarnecki:1997sz}%
  \BibitemOpen
  \bibfield  {author} {\bibinfo {author} {\bibfnamefont {A.}~\bibnamefont
  {Czarnecki}}, \bibinfo {author} {\bibfnamefont {K.}~\bibnamefont {Melnikov}},
  \ and\ \bibinfo {author} {\bibfnamefont {N.}~\bibnamefont {Uraltsev}},\
  }\href {\doibase 10.1103/PhysRevLett.80.3189} {\bibfield  {journal} {\bibinfo
   {journal} {Phys. Rev. Lett.}\ }\textbf {\bibinfo {volume} {80}},\ \bibinfo
  {pages} {3189} (\bibinfo {year} {1998})},\ \Eprint
  {http://arxiv.org/abs/hep-ph/9708372} {arXiv:hep-ph/9708372} \BibitemShut
  {NoStop}%
\bibitem [{\citenamefont {Gambino}(2011)}]{Gambino:2011cq}%
  \BibitemOpen
  \bibfield  {author} {\bibinfo {author} {\bibfnamefont {P.}~\bibnamefont
  {Gambino}},\ }\href {\doibase 10.1007/JHEP09(2011)055} {\bibfield  {journal}
  {\bibinfo  {journal} {JHEP}\ }\textbf {\bibinfo {volume} {09}},\ \bibinfo
  {pages} {055} (\bibinfo {year} {2011})},\ \Eprint
  {http://arxiv.org/abs/1107.3100} {arXiv:1107.3100 [hep-ph]} \BibitemShut
  {NoStop}%
\bibitem [{\citenamefont {Mannel}\ and\ \citenamefont
  {Pivovarov}(2019)}]{Mannel:2019qel}%
  \BibitemOpen
  \bibfield  {author} {\bibinfo {author} {\bibfnamefont {T.}~\bibnamefont
  {Mannel}}\ and\ \bibinfo {author} {\bibfnamefont {A.~A.}\ \bibnamefont
  {Pivovarov}},\ }\href {\doibase 10.1103/PhysRevD.100.093001} {\bibfield
  {journal} {\bibinfo  {journal} {Phys. Rev. D}\ }\textbf {\bibinfo {volume}
  {100}},\ \bibinfo {pages} {093001} (\bibinfo {year} {2019})},\ \Eprint
  {http://arxiv.org/abs/1907.09187} {arXiv:1907.09187 [hep-ph]} \BibitemShut
  {NoStop}%
\bibitem [{\citenamefont {Gambino}(2015)}]{Gambino:2015ima}%
  \BibitemOpen
  \bibfield  {author} {\bibinfo {author} {\bibfnamefont {P.}~\bibnamefont
  {Gambino}},\ }\href {\doibase 10.1142/S0217751X15430022} {\bibfield
  {journal} {\bibinfo  {journal} {Int. J. Mod. Phys. A}\ }\textbf {\bibinfo
  {volume} {30}},\ \bibinfo {pages} {1543002} (\bibinfo {year} {2015})},\
  \Eprint {http://arxiv.org/abs/1501.00314} {arXiv:1501.00314 [hep-ph]}
  \BibitemShut {NoStop}%
\bibitem [{\citenamefont {Amhis}\ \emph {et~al.}(2021)\citenamefont {Amhis}
  \emph {et~al.}}]{Amhis:2019ckw}%
  \BibitemOpen
  \bibfield  {author} {\bibinfo {author} {\bibfnamefont {Y.~S.}\ \bibnamefont
  {Amhis}} \emph {et~al.} (\bibinfo {collaboration} {HFLAV}),\ }\href {\doibase
  10.1140/epjc/s10052-020-8156-7} {\bibfield  {journal} {\bibinfo  {journal}
  {Eur. Phys. J. C}\ }\textbf {\bibinfo {volume} {81}},\ \bibinfo {pages} {226}
  (\bibinfo {year} {2021})},\ \Eprint {http://arxiv.org/abs/1909.12524}
  {arXiv:1909.12524 [hep-ex]} \BibitemShut {NoStop}%
\bibitem [{\citenamefont {Chetyrkin}\ \emph {et~al.}(2000)\citenamefont
  {Chetyrkin}, \citenamefont {Kuhn},\ and\ \citenamefont
  {Steinhauser}}]{Chetyrkin:2000yt}%
  \BibitemOpen
  \bibfield  {author} {\bibinfo {author} {\bibfnamefont {K.~G.}\ \bibnamefont
  {Chetyrkin}}, \bibinfo {author} {\bibfnamefont {J.~H.}\ \bibnamefont {Kuhn}},
  \ and\ \bibinfo {author} {\bibfnamefont {M.}~\bibnamefont {Steinhauser}},\
  }\href {\doibase 10.1016/S0010-4655(00)00155-7} {\bibfield  {journal}
  {\bibinfo  {journal} {Comput. Phys. Commun.}\ }\textbf {\bibinfo {volume}
  {133}},\ \bibinfo {pages} {43} (\bibinfo {year} {2000})},\ \Eprint
  {http://arxiv.org/abs/hep-ph/0004189} {arXiv:hep-ph/0004189} \BibitemShut
  {NoStop}%
\bibitem [{\citenamefont {Gambino}\ and\ \citenamefont
  {Schwanda}(2014)}]{Gambino:2013rza}%
  \BibitemOpen
  \bibfield  {author} {\bibinfo {author} {\bibfnamefont {P.}~\bibnamefont
  {Gambino}}\ and\ \bibinfo {author} {\bibfnamefont {C.}~\bibnamefont
  {Schwanda}},\ }\href {\doibase 10.1103/PhysRevD.89.014022} {\bibfield
  {journal} {\bibinfo  {journal} {Phys. Rev. D}\ }\textbf {\bibinfo {volume}
  {89}},\ \bibinfo {pages} {014022} (\bibinfo {year} {2014})},\ \Eprint
  {http://arxiv.org/abs/1307.4551} {arXiv:1307.4551 [hep-ph]} \BibitemShut
  {NoStop}%
\bibitem [{\citenamefont {Benson}\ \emph {et~al.}(2003)\citenamefont {Benson},
  \citenamefont {Bigi}, \citenamefont {Mannel},\ and\ \citenamefont
  {Uraltsev}}]{Benson:2003kp}%
  \BibitemOpen
  \bibfield  {author} {\bibinfo {author} {\bibfnamefont {D.}~\bibnamefont
  {Benson}}, \bibinfo {author} {\bibfnamefont {I.~I.}\ \bibnamefont {Bigi}},
  \bibinfo {author} {\bibfnamefont {T.}~\bibnamefont {Mannel}}, \ and\ \bibinfo
  {author} {\bibfnamefont {N.}~\bibnamefont {Uraltsev}},\ }\href {\doibase
  10.1016/S0550-3213(03)00452-8} {\bibfield  {journal} {\bibinfo  {journal}
  {Nucl. Phys. B}\ }\textbf {\bibinfo {volume} {665}},\ \bibinfo {pages} {367}
  (\bibinfo {year} {2003})},\ \Eprint {http://arxiv.org/abs/hep-ph/0302262}
  {arXiv:hep-ph/0302262} \BibitemShut {NoStop}%
\bibitem [{\citenamefont {Alberti}\ \emph {et~al.}(2014)\citenamefont
  {Alberti}, \citenamefont {Gambino},\ and\ \citenamefont
  {Nandi}}]{Alberti:2013kxa}%
  \BibitemOpen
  \bibfield  {author} {\bibinfo {author} {\bibfnamefont {A.}~\bibnamefont
  {Alberti}}, \bibinfo {author} {\bibfnamefont {P.}~\bibnamefont {Gambino}}, \
  and\ \bibinfo {author} {\bibfnamefont {S.}~\bibnamefont {Nandi}},\ }\href
  {\doibase 10.1007/JHEP01(2014)147} {\bibfield  {journal} {\bibinfo  {journal}
  {JHEP}\ }\textbf {\bibinfo {volume} {01}},\ \bibinfo {pages} {147} (\bibinfo
  {year} {2014})},\ \Eprint {http://arxiv.org/abs/1311.7381} {arXiv:1311.7381
  [hep-ph]} \BibitemShut {NoStop}%
\bibitem [{\citenamefont {Mannel}\ \emph {et~al.}(2015)\citenamefont {Mannel},
  \citenamefont {Pivovarov},\ and\ \citenamefont {Rosenthal}}]{Mannel:2015jka}%
  \BibitemOpen
  \bibfield  {author} {\bibinfo {author} {\bibfnamefont {T.}~\bibnamefont
  {Mannel}}, \bibinfo {author} {\bibfnamefont {A.~A.}\ \bibnamefont
  {Pivovarov}}, \ and\ \bibinfo {author} {\bibfnamefont {D.}~\bibnamefont
  {Rosenthal}},\ }\href {\doibase 10.1103/PhysRevD.92.054025} {\bibfield
  {journal} {\bibinfo  {journal} {Phys. Rev. D}\ }\textbf {\bibinfo {volume}
  {92}},\ \bibinfo {pages} {054025} (\bibinfo {year} {2015})},\ \Eprint
  {http://arxiv.org/abs/1506.08167} {arXiv:1506.08167 [hep-ph]} \BibitemShut
  {NoStop}%
\bibitem [{\citenamefont {Mannel}\ \emph {et~al.}(2010)\citenamefont {Mannel},
  \citenamefont {Turczyk},\ and\ \citenamefont {Uraltsev}}]{Mannel:2010wj}%
  \BibitemOpen
  \bibfield  {author} {\bibinfo {author} {\bibfnamefont {T.}~\bibnamefont
  {Mannel}}, \bibinfo {author} {\bibfnamefont {S.}~\bibnamefont {Turczyk}}, \
  and\ \bibinfo {author} {\bibfnamefont {N.}~\bibnamefont {Uraltsev}},\ }\href
  {\doibase 10.1007/JHEP11(2010)109} {\bibfield  {journal} {\bibinfo  {journal}
  {JHEP}\ }\textbf {\bibinfo {volume} {11}},\ \bibinfo {pages} {109} (\bibinfo
  {year} {2010})},\ \Eprint {http://arxiv.org/abs/1009.4622} {arXiv:1009.4622
  [hep-ph]} \BibitemShut {NoStop}%
\bibitem [{\citenamefont {Bigi}\ \emph {et~al.}(2007)\citenamefont {Bigi},
  \citenamefont {Uraltsev},\ and\ \citenamefont {Zwicky}}]{Bigi:2005bh}%
  \BibitemOpen
  \bibfield  {author} {\bibinfo {author} {\bibfnamefont {I.~I.}\ \bibnamefont
  {Bigi}}, \bibinfo {author} {\bibfnamefont {N.}~\bibnamefont {Uraltsev}}, \
  and\ \bibinfo {author} {\bibfnamefont {R.}~\bibnamefont {Zwicky}},\ }\href
  {\doibase 10.1140/epjc/s10052-007-0216-8} {\bibfield  {journal} {\bibinfo
  {journal} {Eur. Phys. J. C}\ }\textbf {\bibinfo {volume} {50}},\ \bibinfo
  {pages} {539} (\bibinfo {year} {2007})},\ \Eprint
  {http://arxiv.org/abs/hep-ph/0511158} {arXiv:hep-ph/0511158} \BibitemShut
  {NoStop}%
\bibitem [{\citenamefont {Abudin\'en}\ \emph {et~al.}(2020)\citenamefont
  {Abudin\'en} \emph {et~al.}}]{Abudinen:2020zwm}%
  \BibitemOpen
  \bibfield  {author} {\bibinfo {author} {\bibfnamefont {F.}~\bibnamefont
  {Abudin\'en}} \emph {et~al.} (\bibinfo {collaboration} {Belle-II}),\
  }\href@noop {} {\  (\bibinfo {year} {2020})},\ \Eprint
  {http://arxiv.org/abs/2009.04493} {arXiv:2009.04493 [hep-ex]} \BibitemShut
  {NoStop}%
\bibitem [{\citenamefont {van Tonder}(2021)}]{vanTonder:2021cyp}%
  \BibitemOpen
  \bibfield  {author} {\bibinfo {author} {\bibfnamefont {R.}~\bibnamefont {van
  Tonder}} (\bibinfo {collaboration} {Belle}),\ }in\ \href@noop {} {\emph
  {\bibinfo {booktitle} {{55th Rencontres de Moriond on Electroweak
  Interactions and Unified Theories}}}}\ (\bibinfo {year} {2021})\ \Eprint
  {http://arxiv.org/abs/2105.08001} {arXiv:2105.08001 [hep-ex]} \BibitemShut
  {NoStop}%
\bibitem [{\citenamefont {Melnikov}(2008)}]{Melnikov:2008qs}%
  \BibitemOpen
  \bibfield  {author} {\bibinfo {author} {\bibfnamefont {K.}~\bibnamefont
  {Melnikov}},\ }\href {\doibase 10.1016/j.physletb.2008.07.089} {\bibfield
  {journal} {\bibinfo  {journal} {Phys. Lett. B}\ }\textbf {\bibinfo {volume}
  {666}},\ \bibinfo {pages} {336} (\bibinfo {year} {2008})},\ \Eprint
  {http://arxiv.org/abs/0803.0951} {arXiv:0803.0951 [hep-ph]} \BibitemShut
  {NoStop}%
\bibitem [{\citenamefont {Pak}\ and\ \citenamefont
  {Czarnecki}(2008{\natexlab{a}})}]{Pak:2008qt}%
  \BibitemOpen
  \bibfield  {author} {\bibinfo {author} {\bibfnamefont {A.}~\bibnamefont
  {Pak}}\ and\ \bibinfo {author} {\bibfnamefont {A.}~\bibnamefont
  {Czarnecki}},\ }\href {\doibase 10.1103/PhysRevLett.100.241807} {\bibfield
  {journal} {\bibinfo  {journal} {Phys. Rev. Lett.}\ }\textbf {\bibinfo
  {volume} {100}},\ \bibinfo {pages} {241807} (\bibinfo {year}
  {2008}{\natexlab{a}})},\ \Eprint {http://arxiv.org/abs/0803.0960}
  {arXiv:0803.0960 [hep-ph]} \BibitemShut {NoStop}%
\bibitem [{\citenamefont {Pak}\ and\ \citenamefont
  {Czarnecki}(2008{\natexlab{b}})}]{Pak:2008cp}%
  \BibitemOpen
  \bibfield  {author} {\bibinfo {author} {\bibfnamefont {A.}~\bibnamefont
  {Pak}}\ and\ \bibinfo {author} {\bibfnamefont {A.}~\bibnamefont
  {Czarnecki}},\ }\href {\doibase 10.1103/PhysRevD.78.114015} {\bibfield
  {journal} {\bibinfo  {journal} {Phys. Rev. D}\ }\textbf {\bibinfo {volume}
  {78}},\ \bibinfo {pages} {114015} (\bibinfo {year} {2008}{\natexlab{b}})},\
  \Eprint {http://arxiv.org/abs/0808.3509} {arXiv:0808.3509 [hep-ph]}
  \BibitemShut {NoStop}%
\bibitem [{\citenamefont {Biswas}\ and\ \citenamefont
  {Melnikov}(2010)}]{Biswas:2009rb}%
  \BibitemOpen
  \bibfield  {author} {\bibinfo {author} {\bibfnamefont {S.}~\bibnamefont
  {Biswas}}\ and\ \bibinfo {author} {\bibfnamefont {K.}~\bibnamefont
  {Melnikov}},\ }\href {\doibase 10.1007/JHEP02(2010)089} {\bibfield  {journal}
  {\bibinfo  {journal} {JHEP}\ }\textbf {\bibinfo {volume} {02}},\ \bibinfo
  {pages} {089} (\bibinfo {year} {2010})},\ \Eprint
  {http://arxiv.org/abs/0911.4142} {arXiv:0911.4142 [hep-ph]} \BibitemShut
  {NoStop}%
\bibitem [{\citenamefont {Turczyk}(2016)}]{Turczyk:2016kjf}%
  \BibitemOpen
  \bibfield  {author} {\bibinfo {author} {\bibfnamefont {S.}~\bibnamefont
  {Turczyk}},\ }\href {\doibase 10.1007/JHEP04(2016)131} {\bibfield  {journal}
  {\bibinfo  {journal} {JHEP}\ }\textbf {\bibinfo {volume} {04}},\ \bibinfo
  {pages} {131} (\bibinfo {year} {2016})},\ \Eprint
  {http://arxiv.org/abs/1602.02678} {arXiv:1602.02678 [hep-ph]} \BibitemShut
  {NoStop}%
\bibitem [{\citenamefont {Fael}\ \emph {et~al.}(2019)\citenamefont {Fael},
  \citenamefont {Mannel},\ and\ \citenamefont {Keri~Vos}}]{Fael:2018vsp}%
  \BibitemOpen
  \bibfield  {author} {\bibinfo {author} {\bibfnamefont {M.}~\bibnamefont
  {Fael}}, \bibinfo {author} {\bibfnamefont {T.}~\bibnamefont {Mannel}}, \ and\
  \bibinfo {author} {\bibfnamefont {K.}~\bibnamefont {Keri~Vos}},\ }\href
  {\doibase 10.1007/JHEP02(2019)177} {\bibfield  {journal} {\bibinfo  {journal}
  {JHEP}\ }\textbf {\bibinfo {volume} {02}},\ \bibinfo {pages} {177} (\bibinfo
  {year} {2019})},\ \Eprint {http://arxiv.org/abs/1812.07472} {arXiv:1812.07472
  [hep-ph]} \BibitemShut {NoStop}%
\bibitem [{\citenamefont {Golonka}\ and\ \citenamefont
  {Was}(2006)}]{Golonka:2005pn}%
  \BibitemOpen
  \bibfield  {author} {\bibinfo {author} {\bibfnamefont {P.}~\bibnamefont
  {Golonka}}\ and\ \bibinfo {author} {\bibfnamefont {Z.}~\bibnamefont {Was}},\
  }\href {\doibase 10.1140/epjc/s2005-02396-4} {\bibfield  {journal} {\bibinfo
  {journal} {Eur. Phys. J. C}\ }\textbf {\bibinfo {volume} {45}},\ \bibinfo
  {pages} {97} (\bibinfo {year} {2006})},\ \Eprint
  {http://arxiv.org/abs/hep-ph/0506026} {arXiv:hep-ph/0506026} \BibitemShut
  {NoStop}%
\bibitem [{\citenamefont {Gambino}\ and\ \citenamefont
  {Hashimoto}(2020)}]{Gambino:2020crt}%
  \BibitemOpen
  \bibfield  {author} {\bibinfo {author} {\bibfnamefont {P.}~\bibnamefont
  {Gambino}}\ and\ \bibinfo {author} {\bibfnamefont {S.}~\bibnamefont
  {Hashimoto}},\ }\href {\doibase 10.1103/PhysRevLett.125.032001} {\bibfield
  {journal} {\bibinfo  {journal} {Phys. Rev. Lett.}\ }\textbf {\bibinfo
  {volume} {125}},\ \bibinfo {pages} {032001} (\bibinfo {year} {2020})},\
  \Eprint {http://arxiv.org/abs/2005.13730} {arXiv:2005.13730 [hep-lat]}
  \BibitemShut {NoStop}%
\bibitem [{\citenamefont {Gambino}\ \emph {et~al.}(2017)\citenamefont
  {Gambino}, \citenamefont {Melis},\ and\ \citenamefont
  {Simula}}]{Gambino:2017vkx}%
  \BibitemOpen
  \bibfield  {author} {\bibinfo {author} {\bibfnamefont {P.}~\bibnamefont
  {Gambino}}, \bibinfo {author} {\bibfnamefont {A.}~\bibnamefont {Melis}}, \
  and\ \bibinfo {author} {\bibfnamefont {S.}~\bibnamefont {Simula}},\ }\href
  {\doibase 10.1103/PhysRevD.96.014511} {\bibfield  {journal} {\bibinfo
  {journal} {Phys. Rev. D}\ }\textbf {\bibinfo {volume} {96}},\ \bibinfo
  {pages} {014511} (\bibinfo {year} {2017})},\ \Eprint
  {http://arxiv.org/abs/1704.06105} {arXiv:1704.06105 [hep-lat]} \BibitemShut
  {NoStop}%
\bibitem [{\citenamefont {Jung}\ and\ \citenamefont
  {Straub}(2019)}]{Jung:2018lfu}%
  \BibitemOpen
  \bibfield  {author} {\bibinfo {author} {\bibfnamefont {M.}~\bibnamefont
  {Jung}}\ and\ \bibinfo {author} {\bibfnamefont {D.~M.}\ \bibnamefont
  {Straub}},\ }\href {\doibase 10.1007/JHEP01(2019)009} {\bibfield  {journal}
  {\bibinfo  {journal} {JHEP}\ }\textbf {\bibinfo {volume} {01}},\ \bibinfo
  {pages} {009} (\bibinfo {year} {2019})},\ \Eprint
  {http://arxiv.org/abs/1801.01112} {arXiv:1801.01112 [hep-ph]} \BibitemShut
  {NoStop}%
\bibitem [{\citenamefont {Crivellin}\ and\ \citenamefont
  {Pokorski}(2015)}]{Crivellin:2014zpa}%
  \BibitemOpen
  \bibfield  {author} {\bibinfo {author} {\bibfnamefont {A.}~\bibnamefont
  {Crivellin}}\ and\ \bibinfo {author} {\bibfnamefont {S.}~\bibnamefont
  {Pokorski}},\ }\href {\doibase 10.1103/PhysRevLett.114.011802} {\bibfield
  {journal} {\bibinfo  {journal} {Phys. Rev. Lett.}\ }\textbf {\bibinfo
  {volume} {114}},\ \bibinfo {pages} {011802} (\bibinfo {year} {2015})},\
  \Eprint {http://arxiv.org/abs/1407.1320} {arXiv:1407.1320 [hep-ph]}
  \BibitemShut {NoStop}%
\end{thebibliography}%

\end{document}